Integrating Posture Control in Speech Motor Models: A Parallel-Structured Simulation Approach

Yadong Liu[1], Sidney Fels[2], Arian Shamei[1,3], Najeeb Khan[1], Bryan Gick[1]

1. Department of Linguistics, University of British Columbia
2. Department of Electrical and Computer Engineering, University of British Columbia
3. École de technologie supérieure

Correspondence to: Yadong Liu <yadong@alumni.ubc.ca>

Abstract

Posture is an essential aspect of motor behavior, necessitating continuous muscle activation to counteract gravity. It remains stable under perturbation, aiding in maintaining bodily balance and enabling movement execution. Similarities have been observed between gross body postures and speech postures, such as those involving the jaw, tongue, and lips, which also exhibit resilience to perturbations and assist in equilibrium and movement. Although postural control is a recognized element of human movement and balance, particularly in broader motor skills, it has not been adequately incorporated into existing speech motor control models, which typically concentrate on the gestures or motor commands associated with specific speech movements, overlooking the influence of postural control and gravity. Here we introduce a model that aligns speech posture and movement, using simulations to explore whether speech posture within this framework mirrors the principles of bodily postural control. Our findings indicate that, akin to body posture, speech posture is also robust to perturbation and plays a significant role in maintaining local segment balance and enhancing speech production.

Introduction

Posture is a fundamental motor behavior, and is a precursor to other voluntary movements such as reaching and walking (Ting, 2007). According to Ivanenko and Gurfinkel (2018), posture is described as "tonic muscle contractions against gravity and stabilizing the positions of body segments." Postures can be disrupted yet remain resilient to disturbances, which primarily arise from two main sources: external and internal. External perturbation refers to mechanical perturbations imposed on posture from outside the body's own movements that are usually unexpected, and the corresponding adaptation can be reactive (Rogers and Mille, 2018). For example, when an unexpected weight is added to a bucket held by a participant, increase of activities is observed in the biceps brachii and the thoracic erector spinae muscles after the drop (Moseley et al., 2003). Internal perturbation is imposed by other (often transitory) movements on posture, which can be addressed in part by anticipatory adjustments. This anticipatory posture adjustment (APA) is employed by the central neural system in anticipation of internal perturbation from other movements. APAs play a crucial role in mitigating the destabilizing effects that sudden movements can have on posture, thus preventing potential disturbances in postural equilibrium, as identified by Massion et al. (1999). Further, evidence suggests that the motor cortex contributes to APAs prior to the execution of movement (Yakovenko and Drew, 2009).

Previous literature on posture has focused exclusively on bodily posture and it remains unknown whether the control paradigms that have been proposed for posture of the head, trunk and limbs apply equally to other parts of the body. The oral region, for example, hosts the most complex movement systems in the human body, with such varied and complex behaviors as speaking, facial expression, feeding, swallowing and respiration, all of which may overlap in space and time. Among oral behaviours, speech is one of the most complex but important motor skills. Speech researchers have long discussed the importance of posture. For example, for over a half-century, articulatory setting (AS) has been defined as the oral posture that is required to produce sounds of a language economically and fluently (Honikman, 1964). Öhman (1967: 34) observed based on EMG data that "the articulatory movements of speech are modulations superimposed upon [a] basic posture," suggesting that a certain speech posture acts as a substrate of speech movements.

Nevertheless, until recently, there has been little evidence testing whether speech postures share similar properties with whole body postures. Shamei et al. (2023b) demonstrated that jaw and tongue posture acts against gravity during speech, Easthope et al. (2023) provides evidence that speech posture shares similar cortical control patterns as whole body postures, and Shamei (2024) shows that speech posture stabilizes the vocal tract segments and prepares for upcoming movements in a similar manner to bodily posture. Gick et al. (2017) identifies a lateral tongue bracing posture that is activity maintained throughout English speech. If a speech posture acts as a substrate for movement to be imposed on, the speech posture should also be observed across languages. Liu et al. (2023) showed that the lateral tongue bracing posture is maintained most of the time during speech cross-linguistically in six different languages. Further, the lateral tongue bracing posture has been shown to be robust to external perturbation (Liu et al. 2022). In terms of internal perturbation to posture from movement, Liu et al. (2021) investigated the interaction of smiling postures and lip closing movements during speech and found smiling postural adjustment prior to lip closure.

Thus, while there is abundant recent evidence suggesting that speech postures behave like body postures and they play an important role in speech motor behavior, posture has not traditionally been incorporated into models of speech motor control. This omission might stem from the fact that earlier models predominantly concentrated on the vocal tract's midsagittal plane configuration, while speech postures can occur beyond this plane. Parrell et al. (2019) critically review existing speech motor control models, pointing out that they primarily address kinematic aspects without considering the biomechanics and physical properties of the vocal tract. These models concentrate on the dynamics and timing of speech gestures, sensory feedback, and muscle lengths linked to individual phonemes, positioning them as the primary targets of speech production. Moreover, these models tend to overlook gravity's impact on speech, disregarding postural control. Yet, speech postures perform an anti-gravity function, and evidence suggests that gravity significantly influences speech production (Stone et al., 2007).

Here we present a parallel model (see Fig. 1) of posture and movement control inspired by previous work on body posture (Frank and Earl, 1990; Massion, 1994). This model contains a postural control component (left panel in black) and a movement control component operating in parallel (right panel in gray). Each control component is a hybrid system that combines feedback control and model predictive control (internal model of the plant). Such hybrid structure combines fast pseudo-feedback from the predictive model with precise actual feedback from the feedback model to deliver swift and dependable control. The postural control system is inspired by the control mechanism proposed by Wolpert et al. (1998), which integrates a Smith predictor into a feedback control loop as an observer (For a review on Smith predictors, see Abe and Yamanaka, 2003). The speech movement control mechanism is inspired by the FACTS model proposed by Parrell et al. (2019), which also uses a Smith predictor mechanism to observe a feedback control loop for speech production.

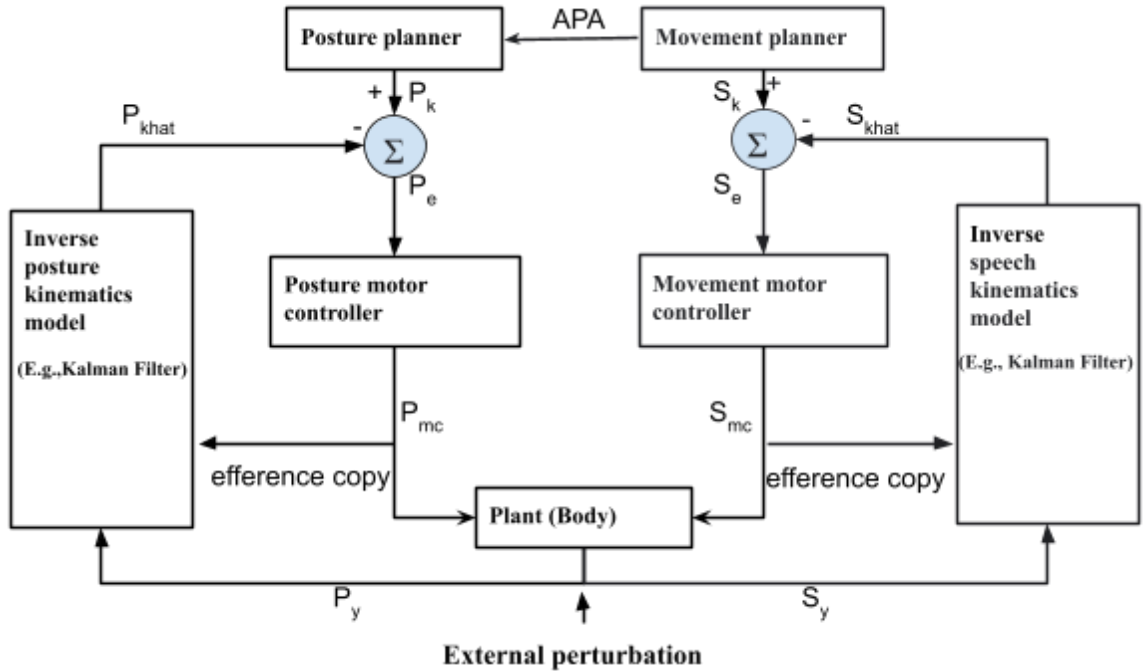

Figure 1. Schematic representation of a parallel speech posture and movement control model. The left part is the posture control component and the right part is the speech movement control component. Both parts use a hybrid control structure that integrates a Kalman Filter that knows the structure of the plant into a feedback control mechanism.

In Fig. 1, The posture planner functions as a neural representation for posture, dispatching a desired kinematic target for posture ($P_k$) to be compared against the current estimated kinematic state of posture ($P_{khat}$), thereby producing a postural kinematic error ($P_e$). This error is channeled to a posture motor controller, which then formulates a motor command ($P_{mc}$) that is relayed to the plant. Concurrently, an efference copy of this motor command is directed to a inverse posture kinematics model, for instance a Kalman filter, serving as an internal model equipped with the plant's characteristics. Upon the plant's execution of the motor command, a resultant motor outcome ($P_y$) is generated and forwarded to the Kalman filter. The filter, using the efference copy, predicts the postural outcome, compares it with the actual motor outcome ($P_y$), and deduces an estimated postural state at the kinematic level ($P_{khat}$). $P_{khat}$ is then aligned against the planner's kinematic target to determine and refine the error, continuing the cycle of posture control.

The movement planner issues a desired speech kinematic target for speech movement ($S_k$) to be matched against the current estimated kinematic state of movement ($S_{khat}$), creating a speech error ($S_e$). This error is forwarded to a speech motor controller, which generates a speech motor command ($S_{mc}$) that is dispatched to the plant. Simultaneously, an efference copy of the motor command is sent to a inverse speech kinematics model (e.g. Kalman filter), acting as an internal model that understands the plant's characteristics. When the plant executes the motor command, a motor outcome ($S_y$) is produced and sent to the Kalman filter. The filter uses the efference copy to predict the movement outcome, compares it with the actual motor outcome ($S_y$), and calculates an estimated speech state at the kinematic level. The estimated speech state of movement is then compared to the kinematic target provided by the speech planner to generate an error, thus perpetuating the cycle of movement control.

The parallel model representing the structure in Figure 1 is simulated using Simulink (MathWorks, 2023). Figure 2 shows the simulated model. Posture planner sends a constant posture target signal of 1 at time 0, and the movement planner sends a transient movement target signal of 5 for 2 seconds at second 5 (raising to 5 from 0 at time 5 and dropping to 0 at time 7). After comparing with the current estimated state, error signals are sent to two PID controllers that have P as 0.5, I as 1, D as 0, and filter coefficient as 100. Integrated control signal is further sent to a plant that uses a transfer function $1/z + 0.5$. Separated copies of control signals are also sent to two Kalman filters, which uses the individual A, B, C, D metrics setting in Simulink (A = -0.5, B = 1, C = 1, D = 0), and uses the G and H metrics setting (G = 0.2, H = 0), and also enables estimated model output y. Estimated states for posture and movement were then compared with their target respectively for the next cycle of simulation.

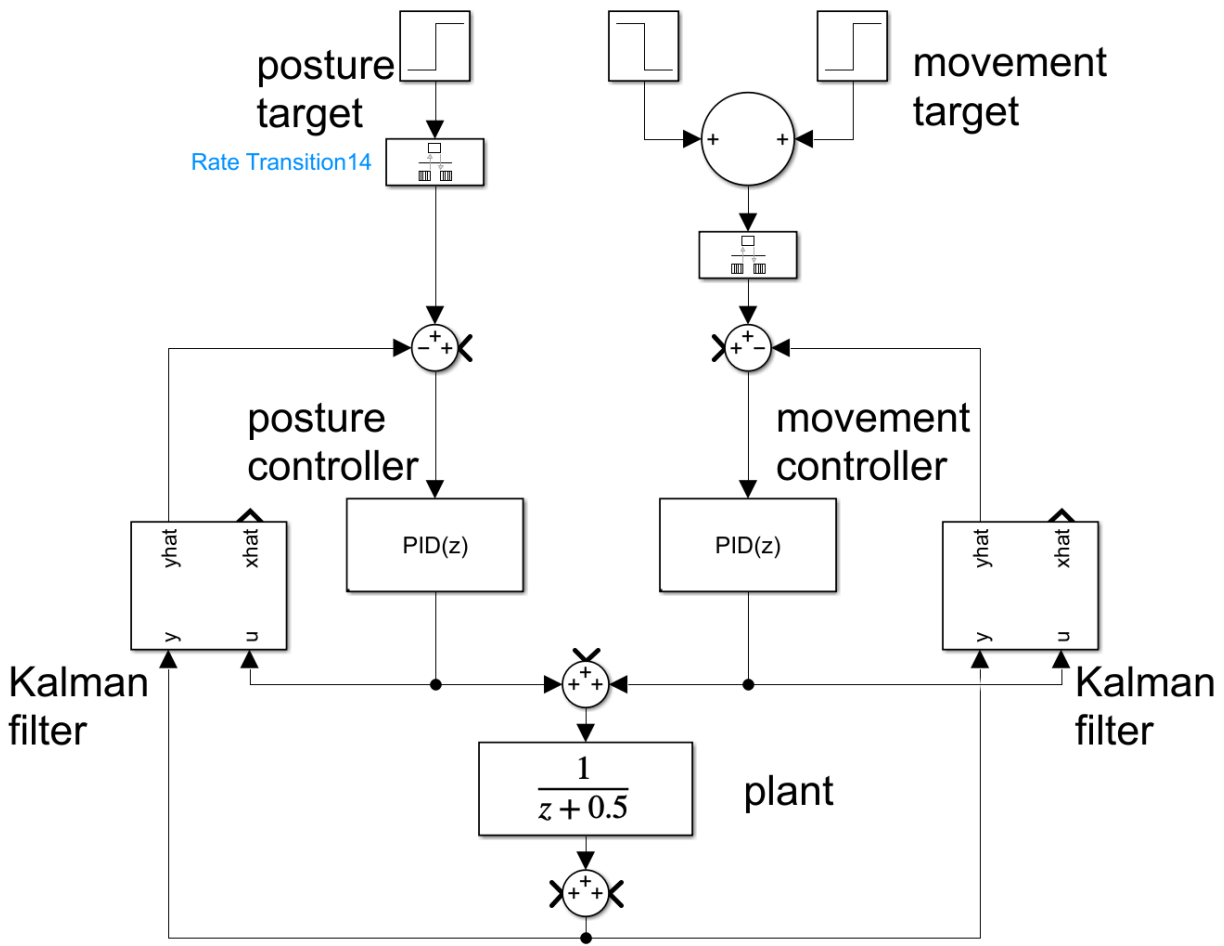

Figure 2. Simulated model structure. The posture planner sends a continuous posture target signal of 1 at time 0, while the movement planner issues a transient movement target signal of 5 for 2 seconds starting at the 5-second mark. An APA (-0.5 from the 3rd to the 5th second) is added to the summation function above the posture controller, and an external perturbation -5 is added to the summation function after the plant. The plant takes the summation of control signals of both controllers and the output represents the summed posture and movement outcome.

Four scenarios are simulated, including a posture-only scenario (posture): a posture target is set as described and movement target is set as 0, a posture with external perturbation scenario (posture with ext. perturbation): same as posture only but adding external perturbation -5 to the summation function after the plant, and a posture and movement co-occurring scenario without APA (posture-movement without APA): posture and movement target is set as described, and a posture and movement co-occurring scenario with APA (posture-movement with APA): same as the previous one but adding an inhibitory APA (-0.5 from the 3rd to the 5th second) to the summation function before postural controller. We hypothesize that speech posture and movement coordination should act similar to body postures and movements. Hence, from our simulations, we predict that in both the posture and posture with ext. perturbation scenarios, the plant will achieve and sustain a robust posture. Additionally, in the posture-movement scenarios, our prediction posits that the plant's outcome will be the cumulative result of the posture target and the movement target. Moreover, we propose that the application of APA will result in reduced errors within the speech movement control, as APA enhances local equilibrium and supports the precise execution of movements.

Figure 3 presents the results of the plant's state from the simulation with input signals showing as colored dashed lines. In the described scenario, the target posture is established at 1, and it is anticipated that the plant will achieve this target (indicated by a red dashed line). Simulation results (refer to Figure 3a) demonstrate that the plant consistently meets this posture target of 1. When an external disturbance is introduced, as seen in the posture with external perturbation scenario, a -5 disturbance impacts the plant's outcome at the 5-second mark. Despite this perturbation, the posture remains robust, and it is predicted to still meet the target of 1 (red dashed line). The corresponding simulation (Figure 3b) reveals that the plant nearly reaches the target, achieving a posture of 0.9975, with the slight deviation attributed to the perturbation. In the scenario without APA, both the posture and movement inputs are set at 1 (red dashed line) and 5 (blue dashed line), respectively. Consequently, the expected outcome during movement execution is 6, as confirmed by the simulation depicted in Figure 3c. However, in the scenario with APA, an inhibitory APA of -0.5 (shown by a brown dashed line from the 3rd to the 5th second) is applied and enhanced movement control accuracy and reduced movement control errors are expected. The simulation (Figure 3d) shows that even with this APA, the plant achieves the projected outcome of 6, although a pre-movement postural perturbation is observed. The results exhibit slight variations: without APA, the outcome is 6.0009; with APA, it drops slightly to 6.0005. Moreover, the movement error decreases from 5 without APA to 4.9997 with APA.

These simulation results indicate that posture maintains its robustness against both external and internal disturbances, with only minor disruptions observed. When comparing the posture-movement scenarios, while the differences in plant outcomes and movement errors with and without APA are minimal, they underscore the role of APA in maintaining equilibrium, reducing movement error and enhancing movement precision, which is also suggested by Bruttini et al. (2016).

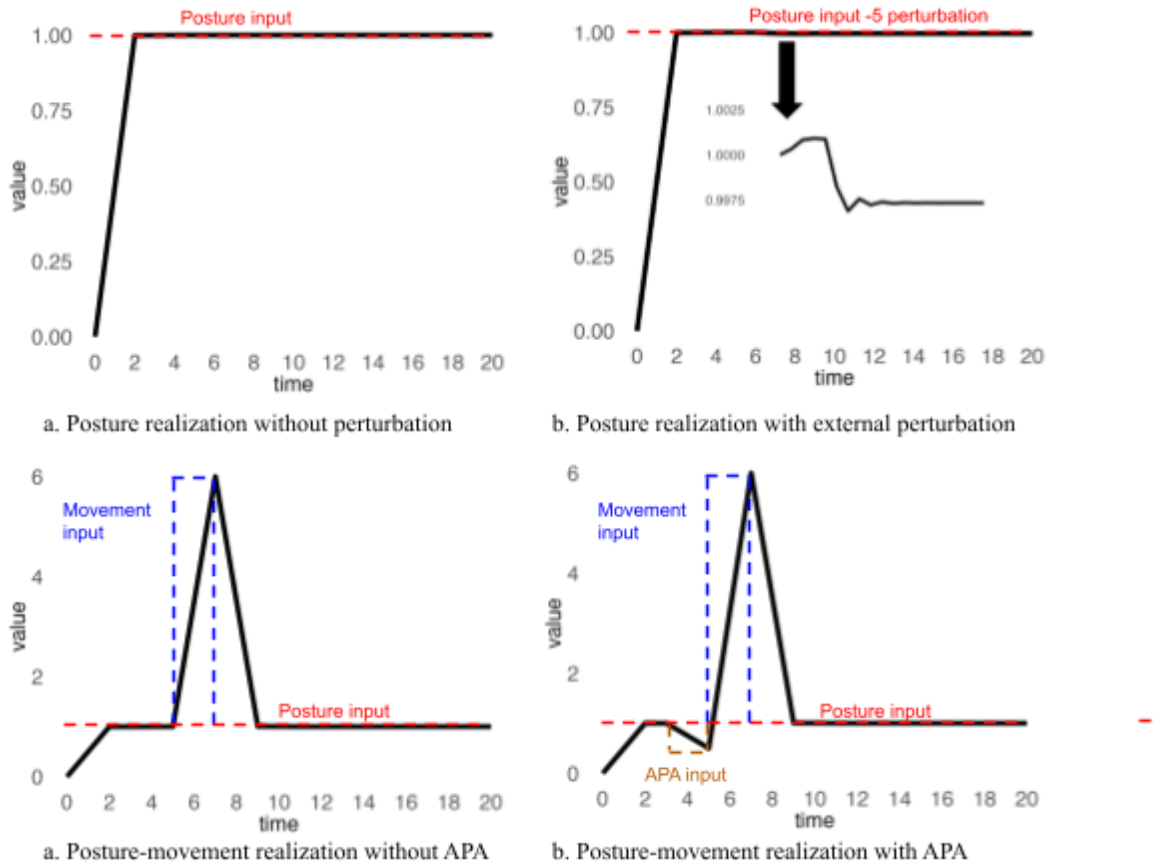

Figure 3. Simulation inputs (in colored dashed lines) and results (in black solid lines). Figure 3a shows simulation input (red dashed line) and results of the posture only scenario. Figure 3b shows simulation input (red dashed line) and results of the posture with ext. Perturbation scenario. Figure 3c shows simulation input (posture in red dashed line and movement in blue dashed line) and results of the posture-movement without APA scenario. Figure 3d shows simulation input (posture in red dashed line, movement in blue dashed line, and APA in brown dashed line) and results of the posture-movement with APA scenario.

Beyond assessing whether speech posture and movement align with whole body postures and gross-motor movements, the proposed parallel speech posture and movement control model also offers the capability to predict speech outcomes influenced by factors other than internal or external perturbations. For example, impaired posture control and posture instability have been noted in patients with neurodegenerative diseases such as Alzheimer's disease (AD) (Fujita et al., 2024) and Parkinson's disease (PD) (Schrag et al., 2002). Correspondingly, in speech motor behaviors, a reduction in vowel space has been observed in patients with AD (Shamei et al., 2023a) and PD (Weismer et al., 2001). However, the specific mechanisms by which neurodegenerative diseases lead to these reduced vowel spaces are not well understood. If neurodegenerative diseases impact overall postural control, then speech postures are likely affected as well. Therefore, employing simulations with the parallel posture and movement speech motor control model can enhance our understanding of the reduced range of articulatory movement observed in patients with neurodegenerative diseases.

In conclusion, though posture plays an important role in speech production, previous speech motor control models have overlooked postural control. By integrating elements from both body posture and movement control and speech motor control, we have developed a novel parallel model for speech posture and movement. Our simulation results reveal that speech postures exhibit similar robustness to bodily postures in the face of perturbations. Furthermore, the application of APAs to speech postures is crucial for maintaining equilibrium and enhancing movement precision, indicating that the principles of postural control are equally vital in the domain of speech.